\documentclass[10pt,aps,twocolumn]{revtex4}
\usepackage{amssymb}
\usepackage{amsmath}
\usepackage{epstopdf}
\usepackage{multirow}
\usepackage[colorlinks=true,linkcolor=red,citecolor=blue]{hyperref}
\begin{document}
\title{Study of  $B \to K_0^*(1430)K^{(*)}$ decays in QCD Factorization Approach}
\author{Ying Li\footnote{Email: liying@ytu.edu.cn}, Hong-Yan Zhang, Ye Xing, Zuo-Hong Li}
\affiliation{Department of Physics, Yantai University, Yantai 264005, China}
\author{Cai-Dian L\"u}
\affiliation{Institute  of  High  Energy  Physics  and  Theoretical Physics Center for Science Facilities,CAS, Beijing 100049, China}
\begin{abstract}
Within the QCD factorization approach, we calculate the branching fractions  and  $CP$ asymmetry parameters of 12  $B \to K_0^*(1430)K^{(*)}$ decay modes under the assumption that the scalar meson $K_0^*(1430)$ is the first excited state or the lowest lying ground state in the quark model. We find that   the decay modes with the scalar meson  emitted,  have large branching fractions due to the enhancement of large chiral factor $r_\chi^{K_0^*}$. The branching fractions of decays with the vector meson  emitted, become much smaller owing to the smaller factor $r_\chi^{K^*}$. Moreover, the annihilation type diagram will induce large uncertainties because of the extra free parameters dealing with  the endpoint singularity. For the pure annihilation type decays, our predictions are smaller than that from PQCD approach by 2-3 orders of magnitudes.  These results will be tested by the ongoing LHCb experiment, forthcoming Belle-II experiment and the proposing circular electron-positron  collider.
\end{abstract}
\date{\today}
\maketitle
\section{Introduction}
Although the quark model has made great success in describing most of the hadronic states, the lowest lying scalar mesons are too light  to fit in the quark model. Two possible scenarios have been proposed on the basis of whether the scalars lower than $1~\mathrm{ GeV}$ belong to the four-quark states or the classical two quark  states. They are controversial for decades \cite{Cheng:2005nb}. In scenario-1 (S1), the scalars such as $\kappa(800)$, $a_0(980)$ and $f_0(980)$ are naively  seen as the lowest lying $q\bar q$ states, and $K_0^*(1430)$, $a_0(1450)$, and $f_0(1500)$ are the  first excited states, correspondingly. In contrast, $K_0^*(1430)$, $a_0(1450)$, $f_0(1500)$ are treated as the the  $q\bar q$ ground states and their first excited states are about $(2.0\sim 2.3)$~GeV in scenario-2 (S2), in which  the lightest scalar mesons are the four-quark bound states.

In the past decade, many $B$ decay modes involving scalar mesons have been reported by BaBar, Belle and large hadron collider-b (LHCb) experiments. This may provide a unique feature to distinguish these two scenarios by the $B$ meson tag \cite{wei}. It is thus hoped that the combination of the precise experimental measurements and the accurate theoretical predictions might provide us valuable information on the nature of scalar mesons.

To achieve this goal, some hadronic $B_q(q=u,d,s)$ decays to scalar mesons have been studied in detail in the framework of the QCD factorization (QCDF) \cite{Cheng:2005nb,Cheng:2013fba, Li:2013aca} and the perturbative QCD approach (PQCD) \cite{PQCD,Liu:2010zg}. Using the QCDF approach, H-Y Cheng {\it et.al}  had calculated the branching fractions and direct $CP$ asymmetries of most decay modes in refs.~\cite{Cheng:2013fba,Li:2013aca}, such as $B \to f_0K$, $B \to K_0^{*}(1430)\phi (\rho) $ and $B \to K_0^{*}(1430)\pi$, where most results accommodate the data with large uncertainties. Very recently, the decays $B \to K_0^{*}(1430) K^{(*)}$ dominated by $b \to d$ penguin operators, have been calculated within the PQCD approach \cite{Liu:2010zg}, however they have not been touched in QCDF frame. To complete, we therefore shall calculate the branching fractions of  $B \to K_0^{*}(1430) K^{(*)}$ decays in QCDF, as well as their $CP$ asymmetry parameters. In the experimental side, these observables are too small to be measured at current experiments, but they are hopeful to be measured in the future experiments such as the updated LHCb experiment, the high luminosity Belle-II experiment and the proposing high energy circular electron-positron collider.

The paper is organized as follows. In Section 2, we give the analytic formula including the effective Hamiltonian, the form factor and all corrections to the amplitudes. Presentation of results and discussions are given in Section 3. At last, we   summarize this work in Section 4.

\section{Analytic Formula}

In this section, we shall start from the weak effective  Hamiltonian responsible for  $B \to K_0^*(1430) K^{(*)}$ decays. In the standard model, it could be written as~\cite{Buchalla}
\begin{multline}\label{eq:eff}
 {\cal H}_{\rm eff}= \frac{G_F}{\sqrt{2}} \biggl[V_{ub}
 V_{ud}^* \left(C_1 O_1^u + C_2 O_2^u  \right) \\ - V_{tb} V_{td}^*\, \big(\sum_{i = 3}^{10}
 C_i O_i  + C_{7\gamma} O_{7\gamma} + C_{8g} O_{8g}\big)\biggl] +
 {\rm h.c.}
\end{multline}
The explicit form of the operators $O_i$ and the corresponding Wilson Coefficients $C_i$ at different  scale $\mu$ could be found in  ref.\cite{Buchalla}. $V_{u(t)b}$ and $V_{u(t)d}$ are the Cabibbo-Kabayashi-Maskawa (CKM) matrix elements. Note that $O_{1,2}$ are tree operators and others $O_{3-10,7\gamma, 8g}$ are penguin ones.

In dealing with the  nonleptonic charmless $B$ decays, the decay amplitude is usually separated into the emission part and the annihilation part in terms of the structure of the topological diagrams. According to the factorization approximation based on the heavy quark limit, the former part could be written as the product of decay constant and form factor. While for the latter one, it is always regarded as being power suppressed. In QCDF  \cite{Beneke:1999br} based on collinear factorization, the contribution of the non-perturbative sector is dominated by the form factors, and the non-factorizable effect in the hadronic matrix elements is controlled by hard gluon exchange. Thus, the hadronic matrix elements of the decay can be written as
\begin{multline}
\langle M_1 M_2|O_i|B\rangle = \sum_{j}F_{j}^{B\rightarrow
M_1 }\int_{0}^{1}dx T_{ij}^{I}(x)\Phi_{M_2}(x) \\+\int_{0}^{1}d\xi\int_{0}^{1}dx\int_{0}^{1}dy
T_{i}^{II}(\xi,x,y)\Phi_{B}(\xi)\Phi_{M_1}(x)\Phi_{M_2}(y),
\end{multline}
where $T_{ij}^{I}$ and $T_{i}^{II}$ denote the perturbative short-distance interactions, which can be calculated perturbatively. $\Phi_{X}(x)~(X=B,M_1,M_2)$ are the universal non-perturbative light-cone distribution amplitudes that can be estimated by the light-cone QCD sum rules. The form factors of $B$ meson to the pseudoscalar ($P$), the vector ($V$) and scalar ($S$) mesons transitions $F^{B\to M_1}$ being a part of the nonperturbative sector of QCD, lack a precise solution. For the $B \to P$ and $B \to V$ transition form factors, we will employ the results of the QCD sum rule method \cite{Ball:2007rt}, since they have been used in calculating $B \to PP, PV$ and $VV$ modes widely. As the form factor of $F^{B\to S}$ as concerned, to the best of our knowledge, a number of approaches had been advocated to calculate them, such as QCD sum rule \cite{Yang:2005bv}, light-cone QCD sum rule \cite{Wang:2008da}, PQCD \cite{Li:2008tk} and covariant light front quark model (cLFQM) \cite{Cheng:2003sm}. In this work, we use the results obtained by  cLFQM \cite{Cheng:2003sm} for keeping consistent with the results of other $B \to S P(V)$ decay modes  \cite{Cheng:2013fba}.

Following the standard procedure of QCDF approach, the emission part of decay amplitude could be written as
\begin{multline}
{\cal A}_S(\overline{B^0}\to M_1M_2) \\
=\frac{G_F}{\sqrt{2}} \sum_{p=u,c}\sum_{i}V_{pb}V_{pd}^{*}a_i^p(\mu)\langle M_1M_2|O_i|B\rangle _{F}.
\end{multline}
$\langle M_{1}M_{2}|O_i|B\rangle_{F} $ is the factorizable matrix element, which can be factorized into a form factor times a decay constant, as stated before. The effective parameters $a_i^p$ can be calculated perturbatively, whose expressions are given by
\begin{eqnarray}  \label{eq:ai}
a_i^p(M_1M_2) &=& \left(C_i+\frac{C_{i\pm1}}{N_c}\right)N_i(M_2) \nonumber\\
&+&\frac{C_{i\pm1}}{ N_c} \frac{C_F\alpha_s}{4\pi}\Big[V_i(M_2)+\frac{4\pi^2}{ N_c}H_i(M_1M_2)\Big]\nonumber \\
&+&P_i^p(M_2),
\end{eqnarray}
with $i=1,\cdots,10$. The upper (lower) signs apply when $i$ is odd (even), $C_F=(N_c^2-1)/(2N_c)$ with $N_c=3$. The quantities $V_i(M_2)$ account for vertex corrections, $H_i(M_1M_2)$ for hard spectator interactions with a hard gluon exchange between the emitted meson and the spectator quark of the $B$ meson and $P_i(M_2)$ for penguin contractions.  Similarly, the annihilation contributions are described by the terms $b_i$,
and $b_{i,{\rm EW}}$, which have the
expressions
\begin{eqnarray}
 b_1 &=& {C_F\over N_c^2}C_1A_1^i, \nonumber\\
 b_2 &=& {C_F\over N_c^2}C_2A_1^i, \nonumber \\
 b_3&=&{C_F\over
 N_c^2}\left[C_3A_1^i+C_5(A_3^i+A_3^f)+N_cC_6A_3^f\right], \nonumber \\
 b_4&=&{C_F\over
 N_c^2}\left[C_4A_1^i+C_6A_2^f\right], \nonumber \\
 b_{\rm 3,EW} &=& {C_F\over
 N_c^2}\left[C_9A_1^{i}+C_7(A_3^{i}+A_3^{f})+N_cC_8A_3^{i}\right],
\nonumber \\
 b_{\rm 4,EW} &=& {C_F\over
 N_c^2}\left[C_{10}A_1^{i}+C_8A_2^{i}\right],
\end{eqnarray}
where the subscripts 1,2,3 of $A_n^{i,f}$ stand for the annihilation amplitudes induced from $(V-A)(V-A)$, $(V-A)(V+A)$ and $(S-P)(S+P)$ operators, respectively. The superscripts $i$ and $f$ refer to gluon emission from the initial and final-state quarks, respectively. It should be stressed that the decays $\overline B^0 \to K_0^{*+}K^{(*)-}$ and $\overline B^0 \to K^{(*)+}K_0^{*-}$ are only induced by the annihilations type diagrams. Hereafter, the order of the $a_i^p(M_1M_2)$ coefficients is dictated by the subscript $M_1M_2$, where $M_2$ is the emitted meson and $M_1$ shares the same spectator quark with the $B$ meson. For the $b_i(M_1M_2)$ of the annihilation, $M_1$ means the one containing an antiquark from the weak vertex, while $M_2$ contains a quark from the weak vertex. The explicit expressions of $V_i$, $H_i $,$P_i $, $b_i$, $b_{i,\rm EW}$ and their inner functions could be found in refs.\cite{Cheng:2005nb,Cheng:2013fba}.

When dealing with the hard-scattering spectator and the weak annihilation contributions, we suffer from infrared endpoint singularities $X=\int_0^1dx/(1-x)$, that cannot be calculated from the first principle in the QCDF approach and only be estimated phenomenologically with a large uncertainty. Following the arguments of ref.\cite{Beneke:1999br}, we also parameterize these kinds of contributions by the complex quantities, $X_H$ and $X_A$ namely,
\begin{eqnarray}
X_{H} &=& \left(1+\rho_{H} e^{i\phi_{H}}\right)\ln\frac{m_B}{\Lambda}, \\
X_{A} &=& \left(1+\rho_{A} e^{i\phi_{A}}\right)\ln\frac{m_B}{\Lambda}~,
\end{eqnarray}
where $\Lambda =0.5 \mathrm{GeV}$.  $\rho_{H}$, $\rho_{A}$ are real parameters, and $\phi_{H}$ and $\phi_{A}$ are free strong phases in the range $[-180^\circ,180^\circ]$. All the above four parameters should be fixed by the experimental data, such as branching fractions and $CP$ asymmetries. In the so-called PQCD approach \cite{Keum:2000ph}, one can eliminate these end-point singularities by keeping all small transverse momenta of gluons and inner quarks.

Including the emission and annihilation contributions, the decay amplitude can be finally given as
\begin{widetext}
\begin{eqnarray} \label{eq:af1}
A(B^- \to K^- K_0^{*0} ) &=&
 \frac{G_F}{\sqrt{2}}\sum_{p=u,c}\lambda_p^{(d)}
 \Bigg\{
 \left(a_4^p -\frac{1}{2}a_{10}^p-(a_6^p-\frac{1}{2}a_8^p)r_\chi^{K_0^*}\right)_{K^-K_0^{*0}} f_{K_0^*}F_0^{B\to K}(m_{K_0^*}^2)(m_B^2-m_K^2)\nonumber \\& &+ f_B f_{K_0^*}f_K\big(b_2\delta_u^p+b_3
 +b_{\rm 3,EW}\big)_{K^-K_0^{*0}} \Bigg\},
\end{eqnarray}
\begin{eqnarray} \label{eq:af2}
A(B^- \to K_0^{*-} K^0 ) &=&
 \frac{G_F}{\sqrt{2}}\sum_{p=u,c}\lambda_p^{(d)}
 \Bigg\{-
 \left(a_4^p -\frac{1}{2}a_{10}^p-(a_6^p-\frac{1}{2}a_8^p)r_\chi^{K}\right)_{K_0^{*-}K^0} f_{K}F_0^{B\to K_0^*}(m_{K}^2)(m_B^2-m_{K_0^*}^2)\nonumber \\& &+ f_Bf_{K_0^*}f_K\big(b_2\delta_u^p+b_3
 +b_{\rm 3,EW}\big)_{K_0^{*-}K^0} \Bigg\},
\end{eqnarray}
\begin{eqnarray} \label{eq:af3}
A(\overline B^0 \to \overline K^0 K_0^{*0} ) &=&
 \frac{G_F}{\sqrt{2}}\sum_{p=u,c}\lambda_p^{(d)}
 \Bigg\{
 \left(a_4^p -\frac{1}{2}a_{10}^p-(a_6^p-\frac{1}{2}a_8^p)r_\chi^{K_0^*}\right)_{\overline K^0K_0^{*0}} f_{K_0^*}F_0^{B\to K}(m_{K_0^*}^2)(m_B^2-m_K^2)\nonumber \\& &+ f_Bf_{K_0^*}f_K\big[\left(b_3+b_4-\frac{1}{2}(
 b_{\rm 3,EW}+b_{\rm 4,EW})\right)_{\overline K^0K_0^{*0}}+\left(b_4
 -\frac{1}{2}b_{\rm 4,EW}\right)_{K_0^{*0}\overline K^0}\big] \Bigg\},
\end{eqnarray}
\begin{eqnarray} \label{eq:af4}
A(\overline B^0 \to \overline K_0^{*0} K^0 ) &=&
 \frac{G_F}{\sqrt{2}}\sum_{p=u,c}\lambda_p^{(d)}
 \Bigg\{
 -\left(a_4^p -\frac{1}{2}a_{10}^p-(a_6^p-\frac{1}{2}a_8^p)r_\chi^{K}\right)_{\overline K_0^{*0}K^0} f_{K}F_0^{B\to K_0^*}(m_{K}^2)(m_B^2-m_{K_0^*}^2)\nonumber \\& &+ f_Bf_{K_0^*}f_K\big[\left(b_3+b_4-\frac{1}{2}(
 b_{\rm 3,EW}+b_{\rm 4,EW})\right)_{\overline K_0^{*0}K^0}+\left(b_4
 -\frac{1}{2}b_{\rm 4,EW}\right)_{K^0\overline K_0^{*0}}\big] \Bigg\},
\end{eqnarray}
\begin{eqnarray} \label{eq:af5}
A(\overline B^0 \to  K_0^{*+} K^- ) &=&
 \frac{G_F}{\sqrt{2}}\sum_{p=u,c}\lambda_p^{(d)}
 \Bigg\{
  f_Bf_{K_0^*}f_K\big[\left(b_1\delta_u^p+b_4
 +b_{\rm 4,EW}\right)_{K_0^{*+} K^-}+\left(b_4
 -\frac{1}{2}b_{\rm 4,EW}\right)_{K^- K_0^{*+} }\big] \Bigg\},
\end{eqnarray}
\begin{eqnarray} \label{eq:af6}
A(\overline B^0 \to K^+ K_0^{*-}) &=&
 \frac{G_F}{\sqrt{2}}\sum_{p=u,c}\lambda_p^{(d)}
 \Bigg\{
  f_Bf_{K_0^*}f_K\big[\left(b_1\delta_u^p+b_4
 +b_{\rm 4,EW}\right)_{K^+ K_0^{*-}}+\left(b_4
 -\frac{1}{2}b_{\rm 4,EW}\right)_{K_0^{*-}K^+ }\big] \Bigg\},
\end{eqnarray}
\begin{eqnarray} \label{eq:af7}
A(B^- \to K^{*-} K_0^{*0} ) &=&
\frac{G_F}{\sqrt{2}}\sum_{p=u,c}\lambda_p^{(d)}
 \Bigg\{
 -\left(a_4^p -\frac{1}{2}a_{10}^p+(a_6^p-\frac{1}{2}a_8^p)r_\chi^{K_0^*}\right)_{K^{*-}K_0^{*0}} 2f_{K_0^*}A_0^{B\to K^*}(m_{K_0^*}^2)m_B p_c\nonumber \\& &-  f_Bf_{K_0^*}f_{K^*}\big(b_2\delta_u^p+b_3
 +b_{\rm 3,EW}\big)_{K^{*-}K_0^{*0}} \Bigg\},
\end{eqnarray}
\begin{eqnarray} \label{eq:af8}
A(B^- \to K_0^{*-} K^{*0} ) &=&
\frac{G_F}{\sqrt{2}}\sum_{p=u,c}\lambda_p^{(d)}
 \Bigg\{
 \left(a_4^p -\frac{1}{2}a_{10}^p-(a_6^p-\frac{1}{2}a_8^p)r_\chi^{K^*}\right)_{K_0^{*-}K^{*0}} 2f_{K^*}F_1^{B\to K_0^*}(m_{K^*}^2)m_B p_c\nonumber \\& &- f_Bf_{K_0^*}f_{K^*}\big(b_2\delta_u^p+b_3
 +b_{\rm 3,EW}\big)_{K_0^{*-}K^{*0}} \Bigg\},
\end{eqnarray}
\begin{eqnarray} \label{eq:af9}
A(\overline B^0 \to \overline K^{*0} K_0^{*0} ) &=&
\frac{G_F}{\sqrt{2}}\sum_{p=u,c}\lambda_p^{(d)}
 \Bigg\{
 -\left(a_4^p -\frac{1}{2}a_{10}^p+(a_6^p-\frac{1}{2}a_8^p)r_\chi^{K_0^*}\right)_{\overline K^{*0}K_0^{*0}} 2f_{K_0^*}A_0^{B\to K^*}(m_{K_0^*}^2)m_B p_c\nonumber  \\& &- f_Bf_{K_0^*}f_{K^*}\big[\left(b_3+b_4-\frac{1}{2}(
 b_{\rm 3,EW}+b_{\rm 4,EW})\right)_{\overline K^{*0}K_0^{*0}}+\left(b_4
 -\frac{1}{2}b_{\rm 4,EW}\right)_{K_0^{*0}\overline K^{*0}}\big]\,,
\end{eqnarray}
\begin{eqnarray} \label{eq:af10}
A(\overline B^0 \to \overline K_0^{*0} K^{*0} ) &=&
\frac{G_F}{\sqrt{2}}\sum_{p=u,c}\lambda_p^{(d)}
 \Bigg\{
 \left(a_4^p -\frac{1}{2}a_{10}^p-(a_6^p-\frac{1}{2}a_8^p)r_\chi^{K^*}\right)
 _{\overline K_0^{*0}K^{*0}} 2f_{K^*}F_1^{B\to K_0^*}(m_{K^*}^2)m_Bp_c\nonumber \\& &- f_Bf_{K_0^*}f_{K^*}\big[\left(b_3+b_4-\frac{1}{2}(
 b_{\rm 3,EW}+b_{\rm 4,EW})\right)_{\overline K_0^{*0}K^{*0}}+\left(b_4
 -\frac{1}{2}b_{\rm 4,EW}\right)_{K^{*0}\overline K_0^{*0}}\big] \Bigg\},
\end{eqnarray}
\begin{eqnarray} \label{eq:af11}
A(\overline B^0 \to  K_0^{*+} K^{*-} ) &=&
 \frac{G_F}{\sqrt{2}}\sum_{p=u,c}\lambda_p^{(d)}
 \Bigg\{
  -f_Bf_{K_0^*}f_K\big[\left(b_1\delta_u^p+b_4
 +b_{\rm 4,EW}\right)_{K_0^{*+} K^{*-}}+\left(b_4
 -\frac{1}{2}b_{\rm 4,EW}\right)_{K^{*-} K_0^{*+} }\big] \Bigg\},
\end{eqnarray}
\begin{eqnarray} \label{eq:af12}
A(\overline B^0 \to K^{*+} K_0^{*-}) &=&
 \frac{G_F}{\sqrt{2}}\sum_{p=u,c}\lambda_p^{(d)}
 \Bigg\{-f_Bf_{K_0^*}f_K\big[\left(b_1\delta_u^p+b_4
 +b_{\rm 4,EW}\right)_{K^{*+} K_0^{*-}}+\left(b_4
 -\frac{1}{2}b_{\rm 4,EW}\right)_{K_0^{*-}K^{*+} }\big] \Bigg\},
\end{eqnarray}
\end{widetext}
where
\begin{eqnarray} \label{eq:r}
 r^K_\chi(\mu)&=&{2m_K^2\over m_b(\mu)[m_u(\mu)+m_s(\mu)]},\nonumber\\
 r^{K^*_0}_\chi(\mu)&=&{2m_{K_0^*}^2\over
 m_b(\mu)[m_q(\mu)-m_s(\mu)]}, \nonumber\\
  r^{K^*}_\chi(\mu) &=&\frac{2 m_{K^*}}{m_b(\mu)}\frac{f^\perp_{K^*}(\mu)}{f_{K^*}}.
\end{eqnarray}

\begin{table*}[htb!]
\centering
\caption{The Parameters of the $K_0^*(1430)$ in different scenarios}
\label{tab:para}
\begin{tabular}{c}\hline\hline
 $\begin{array}{c|ccc|ccc|cccc}
 ~&F_0^{B\to K_0^*}(0) & a &b& F_1^{B\to K_0^*}(0)&a&b
 &\bar f_{K_0^*}\mbox{(MeV)}&B_1&B_3\\\hline
S1& 0.26& 0.44&0.05& 0.26& 1.52& 0.64 &- 300\pm30 &0.39\pm0.05&-0.70\pm0.05 \\
S2& 0.21& 0.44&0.05& 0.21& 1.52& 0.64 &  445\pm50&-0.39\pm0.09&-0.25\pm0.13
 \end{array}$\\
 \hline
\hline
\end{tabular}
\centering
\end{table*}
\begin{table*}[htb]
\centering
\caption{Summary of input parameters}
\label{parameter}
\begin{tabular}{c}\hline\hline
 $\begin{array}{ccccccccc}
 \lambda & A & R_u&\gamma&\Lambda_{\overline{\mathrm{MS}}}^{(f=4)}
 &\tau_{B^0}&\tau_{B^-}&\lambda_B&\alpha_e \\
 0.225& 0.818 &0.376 & 68^\circ& 250 \mbox{MeV}&1.54\mbox{ps}&1.67\mbox{ps}&0.35&1/132
 \end{array}$\\
 \hline
 $\begin{array}{cccccccccc}
f_{B}        &  m_{B}        & f_{K} & f_{K^*} &f_{K^*}^\perp &m_{K}&m_{K^*}&m_{K_0^*}& F_0^{B\to K}&A_0^{B\to K^*} \\
236\mbox{MeV}& 5.36\mbox{GeV}&  131\mbox{MeV}&221\mbox{MeV} &175\mbox{MeV}&0.49\mbox{GeV}&0.89\mbox{GeV}&1.43\mbox{GeV} &0.35 &0.34
 \end{array}$ \\
   \hline\hline
\end{tabular}
\centering
\end{table*}

\section{Numerical Results and Discussion}

To proceed, we shall present numerical results obtained from the  formulas given in the previous section. Firstly, we introduce the adopted parameters in our calculation. Secondly, we give the numerical results and show the theoretical errors due to uncertainty of some parameters. At last, some discussions and comparisons will be added.

The scalar meson, unlike the pseudoscalar one, has two kinds of decay constants, namely the vector decay constant $f_S$ and the scale-dependent scalar decay constant $\bar f_S$,  the definitions of which are give by:
\begin{eqnarray}
&&\langle \overline K_0^*(p)|\bar s\gamma_\mu d|0\rangle=f_{K_0^*}p_\mu,\nonumber \\
&& \langle \overline K_0^*(p)|\bar sd|0\rangle=m_{K_0^*}\bar f_{K_0^*}.
\end{eqnarray}
The two decay constants satisfy
\begin{eqnarray} \label{eq:EOM}
 f_{K_0^*}=\frac{m_s(\mu)-m_d(\mu)}{m_{K_0^*}} \bar f_{K_0^*},
\end{eqnarray}
where $m_s(\mu)$ and $m_d(\mu)$ are the running current quark masses. It should be stressed that the decay constants of $K_0^*(1430)$ have the signs flipped from S1 to S2.  The definition of the form factors for the $B\to S$ transitions are given by \cite{Cheng:2003sm}
\begin{eqnarray} \label{eq:FF}
\langle S(p')|A_\mu|B(p)\rangle &=& -i\Bigg[\left(P_\mu-{m_B^2-m_S^2\over
q^2}\,q_ \mu\right) F_1^{BS}(q^2) \nonumber \\  &+&{m_B^2-m_S^2\over
q^2}q_\mu\,F_0^{BS}(q^2)\Bigg],
\end{eqnarray}
where $P_\mu=(p+p')_\mu$ and $q_\mu=(p-p')_\mu$. In cLFQM , the momentum dependence of the form factor could be parameterized in a di-pole model form \cite{Cheng:2003sm},
\begin{eqnarray}
F(q^2)={F(0)\over 1-a(q^2/m_B^2)+b(q^4/m_B^4)}.
\end{eqnarray}
Together with the decay constants, the parameters $F(0)$, $a$ and $b$ for $B\to S$ transitions in the different scenarios are summarized in Table~\ref{tab:para}.

The twist-2 light-cone distribution amplitude (LCDA) $\Phi_S(x)$ and twist-3 $\Phi_S^s(x)$ and $\Phi_S^\sigma(x)$ for the scalar meson $\overline K_0^*$ made of the quarks $s\bar d$ are given by
 \begin{eqnarray} \label{eq:wfdef}
&&\langle \overline K_0^*(p)|\bar s(z_2)\gamma_\mu d(z_1)|0\rangle= p_\mu \int
 ^1_0 dx e^{i(xp\cdot z_2+\bar xp\cdot z_1)}\Phi_S(x), \nonumber \\
&& \langle\overline K_0^*(p)|\bar s(z_2) d(z_1)|0\rangle=m_{\overline K_0^*}\int
 ^1_0 dx e^{i(xp\cdot z_2+\bar xp\cdot z_1)}\Phi_S^s(x),  \nonumber \\
&&\langle \overline K_0^*(p)|\bar s(z_2)\sigma_{\mu\nu}d(z_1)|0\rangle \nonumber \\&&=  -m_{\overline K_0^*}(p_\mu z_\nu-p_\nu z_\mu)\int  ^1_0 dx e^{i(xp\cdot z_2+\bar xp\cdot z_1)}{\Phi_S^\sigma(x)\over 6},
\end{eqnarray}
with $z=z_2-z_1$, $\bar x=1-x$, and their normalizations are
\begin{eqnarray} \label{eq:wfnor}
 \int_0^1dx \Phi_S(x)&=&f_S,\\ \int_0^1dx \Phi_S^s(x)&=&
 \int_0^1dx \Phi_S^\sigma(x)=\bar f_S.
\end{eqnarray}
The above definitions of LCDAs can be combined into a single matrix element as
\begin{multline} \label{eq:generalwf}
 \langle \overline K_0^*(p)|\bar s_{2\beta}(z_2)d_{1\alpha}(z_1)|0\rangle
= {1\over
 4}\int_0^1 dxe^{i(xp\cdot z_2+\bar xp\cdot z_1)}\\\times\Bigg\{
 p\!\!\!/\Phi_S(x)
 +m_S\left(\Phi_S^s(x) -\sigma_{\mu\nu}p^\mu
 z^\nu{\Phi_S^\sigma(x)\over 6}\right)\Bigg\}_{\alpha\beta}.
 \end{multline}
In general, the twist-2 light-cone distribution amplitude $\Phi_S$
has the form
\begin{multline}  \label{eq:Swf}
\Phi_S(x,\mu)=\bar f_S(\mu)\,6x(1-x)  \\
 \times\left[B_0(\mu)+\sum_{m=1}^\infty
 B_m(\mu)\,C_m^{3/2}(2x-1)\right],
 \end{multline}
where $B_m(\mu)$ are scale-dependence Gegenbauer moments and $C_m^{3/2}(u)$ are the Gegenbauer polynomials. The $B_m$ of different scenarios are also presented in  Table \ref{tab:para}. As for the twist-3 distribution amplitudes, we shall adopt the asymptotic form for simplicity, shown as
\begin{eqnarray} \label{eq:twist3wf}
 \Phi^s_S(x)=\bar f_S , \qquad \Phi^\sigma_S(x)=\bar f_S\, 6x(1-x).
 \end{eqnarray}
For the LCDAs of $K^{(*)}$, we will employ the formulaes obtained from the QCD sum rules \cite{Ball:2004rg,Ball:2006eu}. The other used parameters, such as the CKM elements, the decay constants of the pseudo-scalar and the vector, and the form factors of $B \to K^{(*)}$, are also listed in Table~\ref{parameter} for convenience.

Now, we turn to discuss the numerical results of the concerned decay modes. As we had stated in previous section, when calculating the hard spectator and the annihilation contributions, two endpoint singularities are parameterized phenomenologically by four free parameters, namely $\rho_H,\phi_H$ and $\rho_A,\rho_A $, which  should be determined from the experimental data. In ref.\cite{Cheng:2013fba}, a global fit of $\rho_A$ and $\phi_A$ to the $B\to SP$ data indicates $\rho_A=0.15$ and $\phi_A=82^\circ$ with $\chi^2=8.3$, so in this work for the central values (or ``default" results), $\rho_{A,H}=0.15$ and $\phi_{A,H}=82 ^\circ$ are adopted. In Table~\ref{Table:Br}, we list the calculated branching fractions   of $B \to K_0^*(1430)K^{(*)}$ decays, where the first uncertainty is due to the variations of $B_{1,3}$ and $f_S$, the second comes from the form factors and the strange quark mass, and the last one is induced by weak annihilation and hard spectator interactions in ranges $\rho_{A,H}\in [0,0.3]$ and $\phi_{A,H}\in [0,180^\circ]$.

\begin{table*}[hbt]
\begin{center}
\caption{The branching ratios of  $B \to K_0^*(1430)K^{(*)}$ in the unite  $10^{-7}$.}
\label{Table:Br}
\begin{tabular}{c|c|c}
\hline
\hline
Decay Mode
 & S1
 & S2 \\
\hline
 $\overline B^0 \to\overline K_0^{*0}(1430)K^0$
 & $11.12_{-2.99-3.57-5.05}^{+3.80+7.07+4.91} $
 & $2.39_{-0.85-0.90-2.00}^{+1.20+1.95+2.67}$ \\
  $B^-\to K_0^{*-}(1430)K^0$
 & $5.36_{-1.37-2.36-2.30}^{+1.73+5.13+1.73}$
 & $1.14_{-0.38-0.56-0.92}^{+0.54+1.40+1.17}$\\
  $\overline B^0\to \overline K^0 K_0^{*0}(1430)$
 & $32.27_{-7.83-5.48-6.30}^{+9.41+7.80+5.49}$
 & $40.47_{-10.77-5.38-6.16}^{+13.36+6.09+6.06}$\\
$B^-\to K^-K_0^{*0}(1430)$
 & $23.71_{-5.60-4.61-3.64}^{+6.67+6.73+2.61}$
 & $33.70_{-8.47-4.82-3.94}^{+10.33+5.52+3.37}$\\
$\overline B^0\to K_0^{*+}(1430)K^-$
 & $0.97_{-0.31-0.01-0.44}^{+0.43+0.01+0.63}$
 & $0.58_{-0.29-0.03-0.05}^{+0.45+0.02+0.14}$ \\
 $\overline B^0\to K^+K_0^{*-}(1430)$
 & $8.33_{-2.55-0.10-4.83}^{+3.31+0.07+7.28}$
 & $1.07_{-0.47-0.04-0.97}^{+0.72+0.03+2.27}$ \\
 \hline
 $\overline B^0\to \overline K_0^{*0}(1430)K^{*0}$
 &$0.08_{-0.02-0.04-0.06}^{+0.07+0.07+0.09} $
 &$0.03_{-0.00-0.00-0.02}^{+0.03+0.02+0.03}$\\
 $B^-\to K_0^{*-}(1430)K^{*0}$
 &$0.62_{-0.08-0.18-0.04}^{+0.08+0.21+0.05}$
 &$0.14_{-0.05-0.08-0.03}^{+0.06+0.11+0.03}$ \\
  $\overline B^0\to \overline K^{*0}K_0^{*0}(1430)$
 &$10.86_{-2.24-0.24-0.75}^{+2.54+0.25+0.53}$
 &$20.13_{-4.49-0.49-0.59}^{+5.35+0.50+0.42}$\\
  $B^-\to K^{*-}K_0^{*0}(1430)$
 &$12.71_{-2.72-0.26-1.07}^{+3.14+0.26+1.16}$
 &$21.71_{-4.66-0.53-0.24}^{+5.51+0.54+0.30}$ \\
$\overline B^0\to K_0^{*+}(1430)K^{*-}$
 &$0.33_{-0.10-0.00-0.16}^{+0.14+0.00+0.22}$
 &$0.11_{-0.06-0.00-0.06}^{+0.09+0.00+0.09}$ \\
 $\overline B^0\to K^{*+}K_0^{*-}(1430)$
 &$14.54_{-4.44-0.00-8.31}^{+5.75+0.00+12.48}$
 &$1.84_{-0.83-0.00-1.75}^{+1.26+0.00+3.94}$ \\
\hline
\hline
\end{tabular}
\end{center}
\end{table*}

Here, we shall take $B^-\to K^-K_0^{*0}(1430)$ and $B^-\to K_0^{*-}(1430)K^0$ as examples to illustrate the relative size of each   contribution. The decay amplitude formula of each mode has been presented in eqs.(\ref{eq:af1}) and (\ref{eq:af2}), respectively.  The first decay mode is characterized by $B\to P$ transition with scalar meson emitted; while the second decay mode is characterized by $B\to S$ transition with pseudoscalar meson emitted. Because of the small vector decay constant of scalar meson $K_0^*(1430)$, $f_{K_0^*}F_0^{B\to K}(m_{K_0^*}^2)$ is suppressed relative to
$f_{K}F_0^{B\to K_0^*}(m_{K}^2)$. Therefore, the decay width of the first channel should be suppressed comparing with the second one. The numerical results of these two decay modes are given as below:
\begin{widetext}
\begin{multline} \label{eq:re1}
A(B^- \to K^- K_0^{*0} ) \propto\underbrace{V_{ub}V_{ud}^*\left(0.53+0.14 i\right)+V_{cb}V_{cd}^*\left(0.58+0.08 i\right)}_{emission\,\, diagrams}
 +\underbrace{ V_{ub}V_{ud}^*\left(-0.03 + 0.01 i \right)+V_{cb}V_{cd}^*\left(-0.03+0.01 i\right) }_{annihilation\,\, diagrams},
\end{multline}
\begin{multline} \label{eq:re2}
A(B^- \to K_0^{*-} K^{0} )\propto \underbrace{ V_{ub}V_{ud}^*\left(0.20+0.05 i\right)+V_{cb}V_{cd}^*\left(0.22+0.03 i\right) }_{emission\,\, diagram}
 +\underbrace{ V_{ub}V_{ud}^*\left(0.08 - 0.01 i\right)+V_{cb}V_{cd}^*\left(-0.04-0.01 i\right)}_{annihilation\,\, diagrams}.
\end{multline}
\end{widetext}
From these two equations, it is apparent that the emission diagrams are dominant. However, there is a large enhancement from $O_{6,8}$ operators due to the fact that the chiral factor $r_\chi^{K^*_0}= 12.3$ at $\mu=4.2$ GeV is much larger than $r_\chi^K=1.5$   owing to the larger mass of $K_0^*(1430)$. It follows that $(a_6^p-\frac{1}{2}a_8^p)r_\chi^{K_0^*}$ is much greater than $(a_6^p-\frac{1}{2}a_8^p)r_\chi^{K}$ and $(a_6^p-\frac{1}{2}a_8^p)r_\chi^{K^*}$. This compensate with the suppression of scalar meson decay constant to result in a larger branching ratio of $B^-\to K^-K_0^{*0}(1430)$ with scalar meson emitted.

In the $B \to PP(V)$ and $VV$ decay modes, the weak annihilation contribution is usually expected to be very small because it belongs to the next leading power correction.  However, one can see from Table~\ref{Table:Br} that the uncertainties induced by the weak annihilation are very large, even much larger than the central values in some decay modes, such as $\overline B^0 \to\overline K_0^{*0}(1430)K^0$ and pure annihilation mode $\overline B^0\to K_0^{*+}(1430)K^{*-}$. This phenomenon can be understood as follow, when discussing  the penguin-induced annihilation diagram of $B\to PP$ mode, the decay amplitude is helicity suppressed heavily because the helicity of one of the final-state mesons cannot match with that of its own quarks. On the contrary, this kind of helicity suppression can be alleviated when the scalar meson involved due to the nonzero orbital angular momentum $L_z$ of the scalar meson.

From Table~\ref{Table:Br}, one could notice that our predicted central values for the branching ratios of $\overline K_0^{*0}(1430)K^0$ and  $K_0^{*-}(1430)K^0$ based on S2 are smaller than the results based on S1 by a factor of $4\sim 5$. In contrast, the central values of of $\overline K^0K_0^{*0}(1430)$ and $K^-K_0^{*0}(1430)$ based on S1 are a bit smaller than those of S2. For $\overline K^{*0}K_0^{*0}(1430)$ and $K^{*-}K_0^{*0}(1430)$, the predicted central values of S2 is larger than those of S1 by a factor 2. Although there is difference between the central values of different scenarios, we cannot distinguish two scenarios due to large uncertainties taken by annihilation diagrams, unless one approach were proposed to deal with annihilations effectively in QCDF. In previous studies \cite{Cheng:2013fba}, by comparing the calculated branching ratios of $B \to K_0^* \pi$ and $B \to K_0^* \phi$  with experimental data,  one concluded that the S2 is prefered, i.e. $K^*_0$ is very likely the lowest lying state. If so, the branching fractions of $\overline K^0K_0^{*0}(1430)$ and $K^-K_0^{*0}(1430)$  could be measured by analyzing  the data of $\overline K^0 K^+\pi^-$ and $K^-K^+\pi^-$. Unfortunately, the predicted results of $\overline K_0^{*0}(1430)K^{*0}$ and $K_0^{*-}(1430)K^{*0}$ is too small to be measured with the current data of Belle, but they are hopeful to be measured in the forthcoming Belle-II by analyzing the four-body final states of $K^+K^-\pi^+\pi^-$.

Within the framework of PQCD, X. Liu $et.al$ had calculated these decays \cite{Liu:2010zg}, and both their results and ours are below the experimental upper limits. Comparing our predictions and theirs, we find that for pure annihilations they obtained rather larger branching fractions by keeping the transverse momenta, which are larger than ours by more than two order of magnitude. For decays with emission diagrams, they also got the larger branching ratios with rather large nonfactorizaion diagrams based on both two different scenarios. We hope the future measurements could distinguish two different frameworks.

\begin{table*}[hbt]
\begin{center}
\caption{The direct CP asymmetries ($\%$) of  $B \to K_0^*(1430)K^{(*)}$ }
\label{Table:cp1}
\begin{tabular}{c|c|c}
\hline
\hline
Decay Mode
 & S1
 & S2 \\
\hline
  $B^-\to K_0^{*-}(1430)K^0$
 & $-5.27_{-0.59-0.57-2.82}^{+0.50+0.31+2.59}$
 &$-22.51_{-7.57-9.36-22.86}^{+4.90+5.63+19.61}$\\
 $B^-\to K^-K_0^{*0}(1430) $
 & $-2.06_{-0.65-0.69-6.67}^{+0.60+0.85+5.46}$
 &$-2.60_{-1.76-0.59-5.47}^{+1.61+0.59+3.52}$\\
 \hline
  $B^-\to K_0^{*-}(1430)K^{*0}$
 &$-18.30_{-1.12-0.89-0.99}^{+0.94+0.65+1.27}$
 &$-31.02_{-6.56-7.85-3.80}^{+4.67+4.28+5.06}$\\
 $B^-\to K^{*-}K_0^{*0}(1430)$
 &$5.03_{-1.37-0.17-13.52}^{+1.30+0.17+11.40}$
 &$0.64_{-2.64-0.22-9.22}^{+2.54+0.23+7.57}$\\
\hline
\hline
\end{tabular}
\end{center}
\end{table*}

The $CP$ asymmetries have not been observed in any $B$ decays involving a scalar meson. For the charged decay modes, according the definition,
\begin{eqnarray}
 A_{CP}^{\rm dir}=\frac{A(B^- \to f)-A(B^+ \to \bar f)}{A(B^- \to f)+A(B^+ \to \bar f)},
\end{eqnarray}
the predictions of the direct $CP$ asymmetries   based on two different scenarios of scalar mesons are summarized in Table \ref{Table:cp1}. The resource of this asymmetry is the interference between the emission diagrams and annihilations. The large uncertainties in the latter lead to the large errors in the direct $CP$ asymmetries, as shown in the table.

\begin{table*}[t]
\begin{center}
\caption{The CP-violating parameters $A_{CP}$, $C$, $\Delta C$, $S$  and $\Delta S$  ($\%$) of $B \to K_0^*(1430)K^{(*)}$.}
\label{Table:cp2}
\begin{tabular}{c|c|c|c|c|c|c}
\hline
\hline
Decay Mode
&Scenario
&$A_{CP}$
&$C$
&$\Delta C$
&$S$
&$\Delta S$ \\\hline
\multirow{2}{1.5cm}{$ \overline K_0^{*0}K^0$}
&S1
&$-0.52^{+0.04+0.12+0.08}_{-0.04-0.10-0.13}$
&$ 0.05^{+0.00+0.00+0.01}_{-0.00-0.00-0.01}$
&$-0.03^{+0.00+0.01+0.00}_{-0.00-0.01-0.00}$
&$ 1.00^{+0.00+0.01+0.00}_{-0.00-0.01-0.00}$
&$ 0.00^{+0.00+0.00+0.00}_{-0.00-0.00-0.00}$\\
\cline{2-7}
&S2
&$-0.90^{+0.03+0.06+0.08}_{-0.02-0.04-0.08}$
&$ 0.05^{+0.01+0.01+0.03}_{-0.00-0.01-0.01}$
&$-0.02^{+0.01+0.01+0.02}_{-0.00-0.01-0.00}$
&$ 1.00^{+0.00+0.00+0.00}_{-0.00-0.00-0.00}$
&$ 0.02^{+0.00+0.00+0.00}_{-0.00-0.00-0.00}$\\
\hline
\multirow{2}{1.5cm}{$ \overline K_0^{*0}K^{*0}$}
&S1
&$-0.98^{+0.02+0.01+0.02}_{-0.00-0.01-0.01}$
&$ 0.33^{+0.01+0.00+0.14}_{-0.07-0.03-0.08}$
&$ 0.20^{+0.01+0.00+0.14}_{-0.06-0.03-0.08}$
&$ 0.91^{+0.03+0.02+0.04}_{-0.03-0.02-0.22}$
&$ 0.02^{+0.03+0.03+0.03}_{-0.02-0.02-0.22}$\\
\cline{2-7}
&S2
&$-1.00^{+0.00+0.00+0.00}_{-0.00-0.00-0.00}$
&$ 0.27^{+0.09+0.08+0.21}_{-0.24-0.23-0.33}$
&$ 0.18^{+0.09+0.08+0.20}_{-0.24-0.23-0.33}$
&$ 0.86^{+0.07+0.06+0.07}_{-0.01-0.00-0.22}$
&$-0.13^{+0.07+0.06+0.07}_{-0.01-0.00-0.22}$\\
\hline
\multirow{2}{1.5cm}{$ K_0^{*+}K^{-}$}
&S1
&$ 0.79^{+0.01+0.00+0.02}_{-0.01-0.00-0.05}$
&$-0.04^{+0.01+0.01+0.04}_{-0.01-0.00-0.04}$
&$ 0.06^{+0.01+0.00+0.07}_{-0.02-0.00-0.06}$
&$ 0.04^{+0.02+0.01+0.02}_{-0.02-0.01-0.05}$
&$-0.84^{+0.03+0.01+0.10}_{-0.03-0.01-0.04}$\\\cline{2-7}
&S2
&$ 0.29^{+0.06+0.01+0.35}_{-0.01-0.00-0.98}$
&$-0.02^{+0.03+0.01+0.27}_{-0.02-0.00-0.10}$
&$ 0.25^{+0.04+0.01+0.29}_{-0.03-0.01-0.26}$
&$-0.36^{+0.05+0.04+0.50}_{-0.05-0.02-0.11}$
&$-0.20^{+0.12+0.01+1.02}_{-0.10-0.02-0.35}$\\
\hline
\multirow{2}{1.5cm}{$ K_0^{*+}K^{*-}$}
&S1
&$ 0.96^{+0.00+0.00+0.00}_{-0.00-0.00-0.01}$
&$-0.01^{+0.00+0.00+0.01}_{-0.00-0.00-0.02}$
&$ 0.02^{+0.01+0.01+0.03}_{-0.00-0.00-0.02}$
&$ 0.01^{+0.00+0.00+0.00}_{-0.00-0.00-0.00}$
&$-0.98^{+0.00+0.00+0.02}_{-0.00-0.00-0.01}$\\
\cline{2-7}
&S2
&$ 0.88^{+0.02+0.00+0.05}_{-0.00-0.00-0.66}$
&$ 0.01^{+0.01+0.00+0.14}_{-0.01-0.00-0.02}$
&$ 0.11^{+0.03+0.01+0.23}_{-0.02-0.00-0.11}$
&$ 0.02^{+0.01+0.00+0.48}_{-0.02-0.00-0.02}$
&$-0.91^{+0.02+0.00+0.63}_{-0.02-0.00-0.05}$\\
\hline
\hline
\end{tabular}
\end{center}
\end{table*}

For the neutral decay modes, the situation becomes more complicated due to the fact that both $B^0$ and $\overline B^0$ could decay to the same final states. For illustration, we will take $ \overline B^0(B^0)\to\overline K_0^{*0}(1430)K^0$  as an examples. Four decay amplitudes, $A_f$, $A_{\bar f}$, $\bar A_f$ and $\bar A_{\bar f}$ are denoted as
\begin{eqnarray}
 A_{f}=\langle K^*_0 \overline K^0|B^0\rangle,
 &\,\,\,&
 A_{\bar f}=\langle \overline K^*_0  K^0|B^0\rangle;\nonumber \\
 \bar A_{f}=\langle K^*_0 \overline K^0|\overline B^0\rangle,
 &\,\,\,&
 \bar A_{\bar f}=\langle \overline K^*_0  K^0|\overline B^0\rangle.
\end{eqnarray}
Then, we can write down the $CP$ asymmetry as
\begin{eqnarray}
 A_{CP}=\frac{|A_f|^2+|\bar A_f|^2-|A_{\bar f}|^2-|\bar A_{\bar f}|^2}{|A_f|^2+|\bar A_f|^2+|A_{\bar f}|^2+|\bar A_{\bar f}|^2}.
\end{eqnarray}
Due to $B^0-\overline B^0$ mixing, the four time-dependent decay widths are given by ($f=\overline K_0^{*0}(1430)K^0$)
\begin{eqnarray}\label{eq:shuai}
\Gamma(B^0(t)\to f)&=&e^{-\Gamma t}\frac{1}{2}(|A_f|^2+|\bar{A}_f|^2)\nonumber\\
&&[1+C_f\cos\Delta m t-S_f\sin\Delta m t],
\nonumber\\
\Gamma(\overline{B}^0(t)\to \bar f)&=&e^{-\Gamma t}\frac{1}{2}(|A_{\bar{f}}|^2+|\bar{A}_{\bar{f}}|^2)
\nonumber\\
&&[1-C_{\overline{f}}\cos\Delta m t+S_{\bar{f}}\sin\Delta m t],\nonumber\\
\Gamma(B^0(t)\to \bar f)&=&e^{-\Gamma t}\frac{1}{2}(|A_{\bar{f}}|^2+|\bar{A}_{\bar{f}}|^2)
\nonumber\\
&&[1+C_{\overline{f}}\cos\Delta m t-S_{\bar{f}}\sin\Delta m t],\nonumber\\
\Gamma(\overline{B}^0(t)\to  f)&=&e^{-\Gamma t}\frac{1}{2}(|A_f|^2+|\bar{A}_f|^2)\nonumber\\
&&[1-C_f\cos\Delta m t+S_f
\sin\Delta m t],
\end{eqnarray}
where $\Delta m$ stands for  the mass difference of two mass eigenstate of $B^0/\overline B^0$ meson, and $\Gamma$ for the  average decay width of the $B$ meson. The auxiliary parameters $C_f$ and $S_f$ appearing in above equations are defined by
\begin{eqnarray}
C_f&=&\frac{|A_f|^2-|\bar{A}_f|^2}{|\bar{A}_f|^2+|A_f|^2}, \\ S_f&=&\frac{2\mathrm{Im}(\lambda_f)}{1+|\bar{A}_f/A_f|^2},\\
\lambda_f&=&\frac{V_{tb}V_{td}^*}{V_{tb}^*V_{td}}\frac{\bar{A}_f}{A_f},
\end{eqnarray}
Replacing $f$ by $\bar{f}$, we could obtain the formulaes for $C_{\bar f}$ and $S_{\bar f}$, correspondingly. If the experiment could find the values of $C_f$, $S_f$, $C_{\bar f}$ and $S_{\bar f}$, we thus can obtain four new parameters:
\begin{eqnarray}
C=\frac{1}{2}(C_f+C_{\bar{f}}), &&\Delta C=\frac{1}{2}(C_f-C_{\bar{f}}),\\
S=\frac{1}{2}(S_f+S_{\bar{f}}), &&\Delta S=\frac{1}{2}(S_f-S_{\bar{f}}).
\end{eqnarray}
Physically, $S$ is the mixing-induced $CP$ asymmetry and $C$ is from the direct $CP$ asymmetry, while $\Delta C$ and $\Delta S$ are $CP$-even under $CP$ transformation $\lambda_f\rightarrow 1/\lambda_{\bar{f}}$. In Table.\ref{Table:cp2}, we present our estimations of $A_{CP}$, $C$, $\Delta C$, $S$ and $\Delta S$ for the final states $\overline K_0^{*0}K^0$, $\overline K_0^{*0}K^{*0}$, $ K_0^{*+}K^-$ and $K_0^{*+}K^{*-}$, under  two different scenarios. It should be stressed that in PQCD \cite{Liu:2010zg}, there is no $CP$ asymmetries because of the absence of tree operators. However, in QCDF, by including the penguin contractions ($P_i$), the charming penguin namely, an extra strong phase together with another one from annihilations might lead to the large $CP$ asymmetry, as shown in the table. Again, the large uncertainties of annihilations result in the large error for the pure annihilation decay modes, especially under S2.  We hope these parameters can be measured in future colliders, such as high luminosity Belle-II, LHC-b and even higher energy $e^+e^-$ collider.

\section{Summary}
In this work, we studied the  $B \to K_0^*(1430)K^{(*)}$ decays under two different scalar meson scenarios by using the QCD factorization approach. We calculated the branching fractions and the   $CP$ asymmetry parameters. It is found that the decay modes with the scalar meson  emitted, have large branching fractions due to the enhancement of large chiral factor $r_\chi^{K_0^*}$, with some of the branching fractions   around the corner of Belle II. In contrast, the branching fractions of  decay modes with the vector meson  emitted, are much smaller. Moreover, the annihilation contributions take large uncertainties because of the free parameter from endpoint singularity. For the pure annihilation type decays,  we predicted very small branching fractions, which are  2-3 orders of magnitudes smaller than the results from PQCD.  Some of the decay channels are hopeful to be measured in future colliders, such as Belle-II, LHC-b and even high energy $e^+e^-$ collider.

\section*{Acknowledgments}
This work is partly supported by the National Science Foundation of China (Grants No. 11175151, 11375208 and No. 11235005), and the Program for New Century Excellent Talents in University (NCET) by Ministry of Education of P. R. China (Grant No. NCET-13-0991).


\end{document}